**Effect of annealing on the physical properties of $RuSr_2GdCu_2O_8$ sintered samples and a discussion about magnetic measurements performed by SQUID magnetometer**

M.R. Cimberle, R. Masini, C. Ferdeghini*, C. Artini[+], G. Costa[+]

CNR-IMEM c/o Physics Department via Dodecaneso 33, 16146 Genova

* INFM-LAMIA c/o Physics Department via Dodecaneso 33, 16146 Genova

[+]DICCI Industrial Chemistry Department via Dodecaneso 31, 16146 Genova

**Abstract**

In this paper we present the physical characterization of sintered $RuSr_2GdCu_2O_8$ samples prepared with different procedures. In particular we show how the physical properties are affected by the final step, which is one week annealing in flowing Oxygen, and how extensively the annealing procedure affects the visibility of the superconducting behaviour. Moreover, we analyse the complexity of the d.c. magnetic measurements performed on such type of samples below the superconducting transition and the possibility to decode them.

## I. Introduction

The hybrid Rutheno-cuprate materials belong to the high $T_c$ superconductors family: they were firstly synthesized with the aim to introduce a metallic sheet between the superconducting $CuO_2$ planes[1], but successively they have been extensively studied for the interesting property of becoming superconductor well within the magnetic ordering, appearing at temperatures $T_m$ greater than 100 K. [2-6] Referring in particular to the phase $RuSr_2GdCu_2O_8$ (Ru1212) the magnetic ordering has been observed by neutron diffraction[7,8] and the result is that the Ruthenium spins form a lattice that is antiferromagnetic in three directions, as the Gd one. The same experiment assigned an upper limit of ~ 0.1 $\mu_B$ to the ferromagnetic component. A possible reason for the ferromagnetism of this basically antiferromagnetic compound has been attributed to a symmetry breaking related to the tilting of the Ru-centered oxygen octahedral about an axis in the plane perpendicular to the c-axis. [9] In this case, a Dzyaloshinsky-Moriya interaction may take place to produce a canting of the Ru lattice to which a ferromagnetic moment in the a-b plane is associated. This scenario would allow us to interpret the magnetic measurements that, in fact, show the simultaneous presence of ferromagnetic and anti-ferromagnetic ordering markers. However, the experimental evidence of the above mentioned tilts around an axis perpendicular to the c axis is controversial [3,6]; in addition, it has been noted [10] that low temperature magnetisation data, close to the saturation value, are not consistent with a "small" canting of the anti-ferromagnetic ordering. Finally, the possibility that the ordering is unsaturated anti-ferromagnetism (ferri-magnetism) has been hypothesised, due to mixed valency state of Ruthenium [11]. Therefore, despite a great quantity of work, the question of the magnetic ordering of this compound is still open.

A second puzzling issue in $RuSr_2GdCu_2O_8$ involved its superconducting behaviour. In fact, even when the resistivity behaviour clearly indicates that a sample is superconducting, the magnetic measurements, both FC and ZFC, are ambiguous: sometimes no trace of diamagnetism is seen, sometimes only the ZFC (shielding) signal is detected, rarely the Meissner effect is observed. On the basis of these facts the existence of a true bulk superconductivity has been questioned for a long time and has not been generally agreed upon up to now [12,13]. The situation is made more complicated by the possibility of experimental artefacts related to the measurement technique: in particular, we refer to magnetic measurements performed by commercial SQUID d.c. magnetometer (widely used in different laboratories), where a series of problems related to the field dis-homogeneity and to the presence of different and opposite magnetic signals together with the need to use small magnetic fields make the interpretation of the magnetic measurements very complex.

In this paper we present and discuss resistive and magnetic measurements of as prepared and annealed samples in correlation with their different preparation treatment. Furthermore, we discuss the interpretation of magnetic measurements and possible experimental artefacts in detail.

## II. Sample preparation

Like in the synthesis procedure previously reported [2,14-16], Ru-1212 was synthesized by solid state reaction of stoichiometric quantities of high purity $RuO_2$, $Gd_2O_3$, CuO and $SrCO_3$.

Required amounts of these materials were ground, preheated at 960°C in air for 12 h in order to promote the decomposition of $SrCO_3$, then reground and reacted at 1010°C in flowing nitrogen for 12 h to obtain precursor materials ($Sr_2GdRuO_6$ and $Cu_2O$) and minimize the formation of $SrRuO_3$. Successively, the resulting samples were pulverized, pressed into pellets and calcined in seven successive steps from 1020°C up to 1070°C, each time for 20 h in flowing oxygen with intermediate grindings. In each reaction step, the samples were quickly cooled down to room temperature. The samples obtained after last thermal treatment are hereafter referred to as "as prepared” samples. Successively, bar shaped pellets from the same “as prepared” batch were annealed in flowing oxygen at 1060°C for 170 days and slowly cooled down to room temperature to obtain the “annealed” samples. All the samples have been synthesized in two batches to test for reproducibility.

The crystal structure was determined by powder X-ray diffraction technique using Cu $K_\alpha$ radiation. Lattice parameters were refined using the Bragg peaks over the θ range. Resistivity measurements were performed by the standard four probe technique with 1 mA current in a closed-cycle helium cryostat in the 13-300 K temperature range and dc magnetization measurements were performed using a Quantum Design SQUID magnetometer.

## III. Results and discussion

### III-1 X-ray diffraction measurements

XRD data indicate that all the as prepared and annealed samples are single phase with tetragonal structure (space group *P4/mmm*) and lattice parameters listed in Table 1.

Table 1 : Structural data of the as prepared and annealed Ru1212 samples

| Sample | a ( Å ) | c ( Å ) | V (Å$^3$) |
|---|---|---|---|
| as prepared | 3.840 | 11.560 | 170.46 |
| annealed | 3.838 | 11.579 | 170.56 |

The annealing produces a variation of the c axis that increases by 0.16 %. In analogy with what happens in YBCO(123) samples, we suggest that this variation can be due to a small oxygen variation and/or re-ordering. SEM analysis on the two samples indicates that no spurious phase is present. A sponge structure is seen in both the samples and the annealing increases both the grain and the void dimension. Roughly the mean grain size varies from about 5 up to about 10 μm, from as prepared and annealed sample.

### III-2 Resistivity measurements

Resistivity measurements and their derivatives for as prepared and annealed samples are shown in Fig.1. The behaviour is the expected one[5,17,18]: from high temperature a linear metallic-type decrease is seen with a small downturn at the temperature of the magnetic ordering $T_m \cong 130K$. Below T=100K a semiconducting upturn is observed down to about T=44K, where the superconducting transition starts, followed by a broad fall to zero that is reached at T= 23K and 33K for as prepared and annealed samples, respectively. This description is common to the two samples. The differences lie: i) in the resistivity values, which are lower in the annealed sample, ii) in the size of the semiconducting upturn and in the width of the superconducting transition, both reduced in the annealed sample, iii) in the onset of the superconducting transition, which are T = 44K and T = 46K for the as prepared and annealed samples, respectively. The details of the superconducting transition are well displayed by the temperature derivative of resistivity in the inset of Fig.1. Two peaks are clearly present in the as prepared sample at 38K and 28K, respectively. For the annealed sample the peaks are nearly overlapping and the first looks like a shoulder: they are at T=40 K and 36K . We point out that this behaviour, already observed in such compound and attributed to the granularity[5,17], is the standard one in high $T_c$ superconductors where “intragrain” and “intergrain” properties give rise to a “double step” transition more or less evident depending on the sample characteristic and the values of the physical parameters involved[19,20]. It may be led back to the effect of the annealing, which increases structure ordering, grains dimension and quality of the weak-link connection between grains and, as a consequence, the zero resistivity temperature. Regarding the slight temperature variation of the superconducting onset, we think it may be related to the oxygen variation and/or re-ordination that we have inferred from the c-axis variation.

### III-3 Magnetic measurements

The magnetic characterization is a key tool to observe both the superconducting and magnetic behaviour of these samples, but it turns out to be a non trivial task.

As already mentioned, the standard diamagnetic signals, both in the FC and ZFC mode, are not always seen in such compound[12,15,21,22]: what is more often observed is the shielding signal, rarely the diamagnetism related to the Meissner effect. Both signals are quickly removed by the application of even a small external magnetic field (few tenth of Gauss). In contrast, even when in the magnetic measurements there is no trace of superconducting behaviour, it may be observed resistively and the application of even high external magnetic fields (up to some Tesla) does not destroy it[23,24]. This is true also for our samples that exhibit zero resistivity up to 2 T, the maximum field at which they were tested[25].

Anomalous behaviours are often measured in the d.c. magnetization of the samples when they enter the superconducting state. Such behaviours, observed since the first magnetic measurements on this material, have been recently reviewed[26]. Briefly, they may be described as follows: at the superconducting transition temperature, when a diamagnetic signal is expected, a peak is seen in the magnetization before it decreases. But also the "reversed" effect has been observed, i.e. the expected decrease at the transition is followed by a sudden increase giving rise to a minimum. Moreover, we will show that both FC and ZFC signals are not reversed by reversing the applied magnetic field.

Different reasons must be considered to understand such phenomenology.

First of all, the simultaneous presence of magnetic and superconducting ordering implies consequences both on the physics and on the experimental measuring technique.

As noted in[9], the magnetization of rutheno-cuprates materials contains signals arising from different contributions: the RE paramagnetic spin lattice (Gd in most cases), the Ru spin lattice and, finally, the diamagnetic signal related to the superconducting behaviour. Such coexistent and opposite signals make the magnetic measurement unsuitable for the observation of the superconductivity: in fact, the signals related to superconductivity and magnetic ordering cannot be separated. Moreover, it is clear that the application of an external magnetic field exalts the magnetic signal and depresses the superconducting one, destroying very quickly its visibility. In the light of these considerations we can understand why the superconducting behaviour is often observed resistively but not magnetically: its visibility depends on the competition between two opposite magnetic signals, and the superconducting must compete with the magnetic one. "More superconductivity" may be related to many factors: the amount of superconducting phase inside the sample, but also the quality of the connection between the grains that makes smaller or larger the shielded or flux free volume and, as a consequence, smaller or larger the related magnetic signals.

Moreover, μ-SR[5] and EPR[27] measurements indicate the presence of an internal field that, at low temperature, may reach hundreds of Gauss and may give rise to a spontaneous vortex phase (SVP) in the temperature range where it exceeds the first critical field $H_{c1}(T)$. [28] In a type II superconductor at $H>H_{c1}$ the complete Meissner effect is never observed because of the presence inside the material of "pinning centres" that block the flux lines and prevent their expulsion. For this reason the FC signal may be very small and its difference from the ZFC signal is an indication of the critical current density that a sample can carry. A vast literature related to high $T_c$ superconductors illustrates unambiguously this issue. [29,30]

Finally, we recall that the samples are generally granular and, therefore, all the problems related to the granular behaviour of HTSC and, in general, to the distinction between intra-granular and inter-granular properties must be taken into account; in particular, the grain size and the quality of the weak links at grain boundaries may influence deeply the phenomenology, enhancing or depressing the superconducting signal in comparison with the magnetic one.

In the following we will present some experimental results: we will start to show measurements where a "standard" behaviour is observed at the superconducting transition, then we will discuss the underlying problems and how to deal with them.

In Fig. 2 the FC and ZFC susceptibility versus temperature measurements for as prepared and annealed samples are compared. The applied magnetic field was 1 G.

At the magnetic transition the two samples exhibit similar behaviours. A sudden onset of a spontaneous magnetic moment, related to a ferromagnetic component, appears in the 130K-140K range. Both the samples have a cuspid in the ZFC curve at $T_m \cong 132K$ while the ZFC and FC curves merge fifteen degrees above this temperature.

It is noteworthy to observe the different behaviours exhibited at the superconducting transition. The diamagnetic signal is much more marked for the annealed sample than for the as prepared one in both FC and ZFC measurements. In ZFC measurements a clear decrease is seen at 10K and 22K for as prepared and annealed samples, respectively, but a first decrease is observed at higher temperature (about 40K) in the annealed sample. This fact is better seen in the temperature derivative of susceptibility (inset of Fig. 2) that shows a double step behaviour, in agreement with what observed in resistivity measurements. Also the magnitude of shielding and flux expulsion are quite different and much greater for the annealed sample. At the lowest temperature of 5K the shielded volumes are about 100% and 10% for as prepared and annealed samples respectively.

In Fig. 3 ZFC and FC data for the annealed sample are presented at magnetic fields ranging from $\mu_0 H$ =3.5 G up to 500 G. At $\mu_0 H$= 3.5 Gauss the ZFC signal corresponds to a shielding of 64% of the total volume while the FC signal corresponds to a flux expulsion from 12% of the volume. A two step transition similar to that observed in Fig.2 occurs. By increasing the field the ZFC transition broadens greatly, as expected in a granular system. At $\mu_0 H$ =50 G a plateau is seen in ZFC curve down to the lowest temperature instead of the diamagnetic decrease and a slight increase in FC curve. A field of 500 G destroys the magnetic visibility of the superconductivity. The general behaviour of susceptibility and its field dependence suggests that a bulk superconductivity is observed.

At the magnetic transition the increase of the field shifts the merging of the ZFC and FC curve to lower temperature. At 500 Gauss the merging temperature is T=132K and the cuspid is completely suppressed.

We point out that at the superconducting transition the SQUID magnetometer indicates a worsening of the quality of the measurement through the regression factor and the answer function that tends to lose its symmetry. This fact does not depend on the magnitude of the signal, which does not change substantially, but on different problems that will be mentioned now.

As a consequence of the complex magnetic signal present in this type of sample, the basic condition required by the SQUID d.c. magnetometer, i.e. a homogeneous magnetic moment that can be assimilated to a point-like dipole, is not fulfilled below the transition temperature. This fact degrades the quality of the measurements. Moreover, as firstly observed in[14], during the measurements the sample is moved for a length that is usually of few centimetres, so that it travels in a non uniform magnetic field that makes it follow a small hysteresis loop. If the value of the moment is not constant during the scan, an asymmetric scan wave form will be observed and also in this case the quality of the measurement will drastically worsen[31]. This last fact (the movement of the sample) has been ascertained as the origin of the peak, or rather, one of the origins of the peak in[16]. The combination of all these facts produces a bad regression of the measurements in the superconducting region that is itself a precise indication of the onset of the significant superconducting signal.

Finally, we point out a last non negligible experimental problem. In the superconducting coil of any experimental set-up a remanent magnetic field is always present. It may be minimized by applying coercive fields of decreasing intensity. In such a way the field is zeroed but for few Gauss, which may be further reduced in the central point of the magnet by applying a small counter-field. However, a very small residual field survives and turns out to be of the order of fractions of Gauss. In our case we verified that it is of few tenth of Gauss by normalizing high temperature susceptibility data taken in this field with other taken in a known field. As a consequence a real ZFC measurement cannot be made and, since the FC magnetic moment is about one order of magnitude greater than the ZFC, also a residual field of fractions of Gauss may give a considerable magnetic signal in particular if, as usually made, small magnetic fields are applied in order to enhance the visibility of the superconducting behaviour. Therefore, when a ZFC measurements is started, the sample is already in a well defined magnetization state whose polarity depends on the residual field polarity, and we have verified that the induced signal is roughly summed to the pre-existent one. It is now clear that the direction of the applied magnetic field is not indifferent because the induced signal may be concordant or opposite in sign to the pre-existent one (that will be called in the following background signal). Two examples are reported in Fig. 4 a) and b). Fig 4 a) shows ZFC data at $\mu_0H_{ext}$=1G in the as prepared sample together with the background signal. The field has been applied in the two opposite directions with respect to the residual field and the results are completely different: with the field concordant the background signal hinders the visibility of the cusp at the magnetic transition, but at the superconducting one the diamagnetic shift is very clearly seen. When the field is opposite, the cusp is quite visible, the background signal draws the ZFC curve towards positive values and, at the superconducting transition, a decrease in the magnetization, not compatible with a diamagnetic shift of the measured curve, but memory of the Meissner effect of the background signal, is observed. If the background signal is subtracted to the two ZFC curves, they become substantially symmetric and by changing a sign they collapse into each other (see Fig.4 a). A difference remains at the superconducting transition where the diamagnetic shift is much smaller for the signal measured with opposite field; however, we point out that here the quality of the measurements is very bad, not only for the aforesaid reasons but also because the signal is approaching zero and the first points must be considered to be not much reliable. In Fig. 4 b) the same is shown for the annealed sample. The behaviour is substantially the same, but clearer at the superconducting transition that is much more visible in this sample. We point out that also in this case the correction is convincing**:** it makes the two shielding signals much more similar**,** and substantially removes the peak present at the superconducting transition in the ZFC curve measured with a field opposite to the residual field. We point out that all the ZFC measurements presented before (Figs 2 and 3) have been corrected for the background signal. As obvious, the correction becomes more and more negligible as the applied magnetic field is increased, but the effect of the background field is still visible at $\mu_0 H_{ext}$ =10G, the maximum field at which we performed measurements inverting the field direction.

The addictivity of the signals makes sense in the ZFC measurements, not in the FC ones. In this case, if a residual field is present, we expect that the application of a magnetic field in its direction or in the opposite one only means to apply different magnetic fields. Therefore, two FC curves measured with applied field in opposite directions should be symmetric but should have slightly different values. We have verified that this is true in general, but it becomes false when the sample enters the superconducting state. With a field applied in the direction of the residual field a diamagnetic shift is observed, but not with an applied field in opposite direction. A clear example of this behaviour is reported in Fig.5 that shows FC measurements at $\mu_0 H_{ext}$=1G applied in the two directions for the annealed sample. The ZFC measurements are the those of Fig 4 b) corrected. As one can see, in the magnetic region the curves are different in magnitude but symmetric: they indicate that the background field is, as expected, of few tenth of Gauss (but important compared to the applied field of 1 G). Whatever direction the field is applied, the magnetic signal behaves exactly in the same way at the superconducting transition, i.e. while in one case (field concordant) it exhibits a decrease in magnetization, in the other (field opposite) it shows an increase in magnetization. When the applied magnetic field is increased the asymmetry between the two measurements tends to null, but the effect observed at the superconducting transition persists up to 10 G (the maximum field at which measurements in the two field directions have been made).

We have looked for some reasonable explanation of this fact without result: we can only mention this problem and suggest taking it into account.

Finally, in Fig.6 we present the M(H) curves measured at T=2K for as prepared and annealed samples. The curves present a bump below 1.5T that we attribute to a spin flop of the Gd lattice (the bump disappears at temperatures higher than 2.5K that is its anti-ferromagnetic ordering temperature). The curve relevant to the annealed sample is higher than that of the as prepared one, and at the maximum field of 5.4 T they tend to saturate to two values differing by about 6%. The same trend may be seen in the inset where hysteresis loops at T=5K are presented. For the annealed sample the hysteresys loop shows not only higher slope but also larger residual magnetisation (0.1295 and 0.1525 $\mu_B$ for as prepared and annealed samples, respectively). This difference, clearly not ascribable to experimental errors, must involve some structural variation. We suggest that, in line with the small increase in the c-axis and in $T_c$, it may be due to a small oxygen variation and/or re-ordination during the annealing that can change the valence state of Ru, and, as a consequence, the magnetic moment.

**IV. Conclusions**

In this paper we have presented a characterization of sintered $RuSr_2Gd\ Cu_2O_8$ samples differing for the final annealing step. The main results are the following: the annealing changes the sample microstructure increasing grains dimension and their connection. Moreover, we think it produces a small variation in the oxygen content and/or ordering (undetectable with standard TGA analysis) to which we attribute the variations of c-axis length, critical temperature and magnetic moment we have detected.

The different granularity leads to different resistive and magnetic behaviours. It plays a role in the resistive, but also in magnetic measurements, increasing the length scale of the shielding or flux expulsion.

The magnetic measurements entail many problems: we have tried to review them, analysing the effect of the residual magnetic field in particular. We point out the lack of symmetry of the signals when the magnetic field is reversed and how it is possible to take care of this effect (at least to the first order) in ZFC measurements by considering the presence of the residual magnetic field. On the contrary, in the FC measurements we have not found a reasonable explanation up to now. Work is in progress on this problem.

Figure Captions

Fig.1 Resistivity versus temperature for as prepared (triangles) and annealed (circles) samples. In the inset dρ/dT as obtained from the data of Fig.1 is shown

Fig. 2 ZFC and FC susceptibility versus temperature for the as prepared and annealed samples. The applied magnetic field is $\mu_0 H$=1Gauss. Open symbols indicate ZFC measurements, full symbols indicate FC measurements. In the inset the derivatives of the ZFC susceptibilities versus temperature magnified in the region around the superconducting transition are shown.

Fig.3 ZFC and FC susceptibility versus temperature for the annealed sample. The applied magnetic fields vary from $\mu_0 H$=3.5 Gauss up to $\mu_0 H$=500 G.

Fig.4 ZFC magnetization versus temperature at $\mu_0 H$=1G applied in two opposite directions. Also the signal measured in the residual field (background signal) is shown and the ZFC magnetizations corrected by subtraction of the background signal: a) as prepared sample b) annealed sample.

Fig.5 FC magnetization versus temperature for the annealed sample. The applied field is 1G in the two opposite directions. Also ZFC magnetizations for the same sample, corrected as indicated in the text, are shown.

Fig.6 Magnetization versus field at T=2K for as prepared and annealed samples. In the inset hysteresis loops at T=5K are shown.

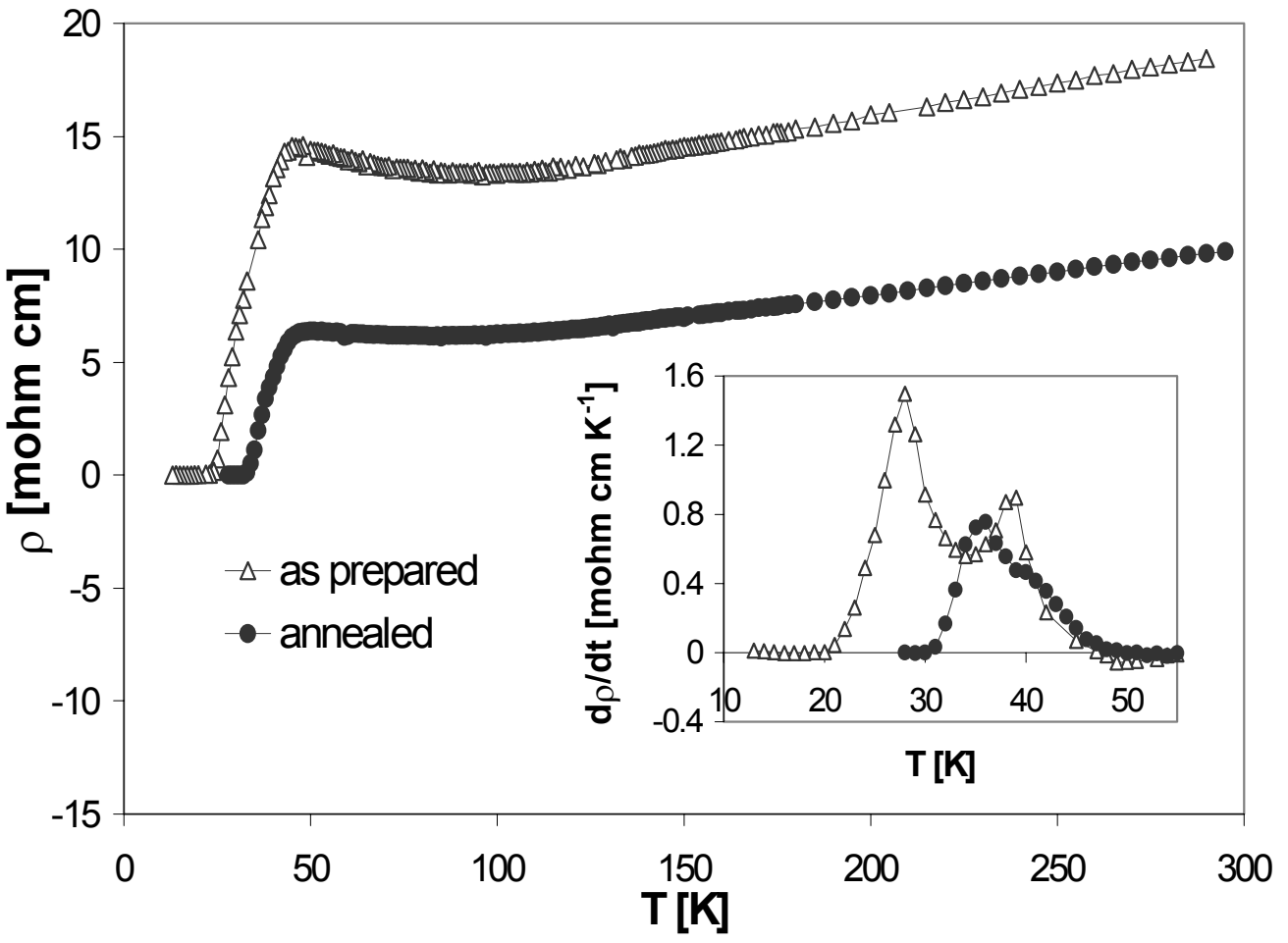

Figure 1

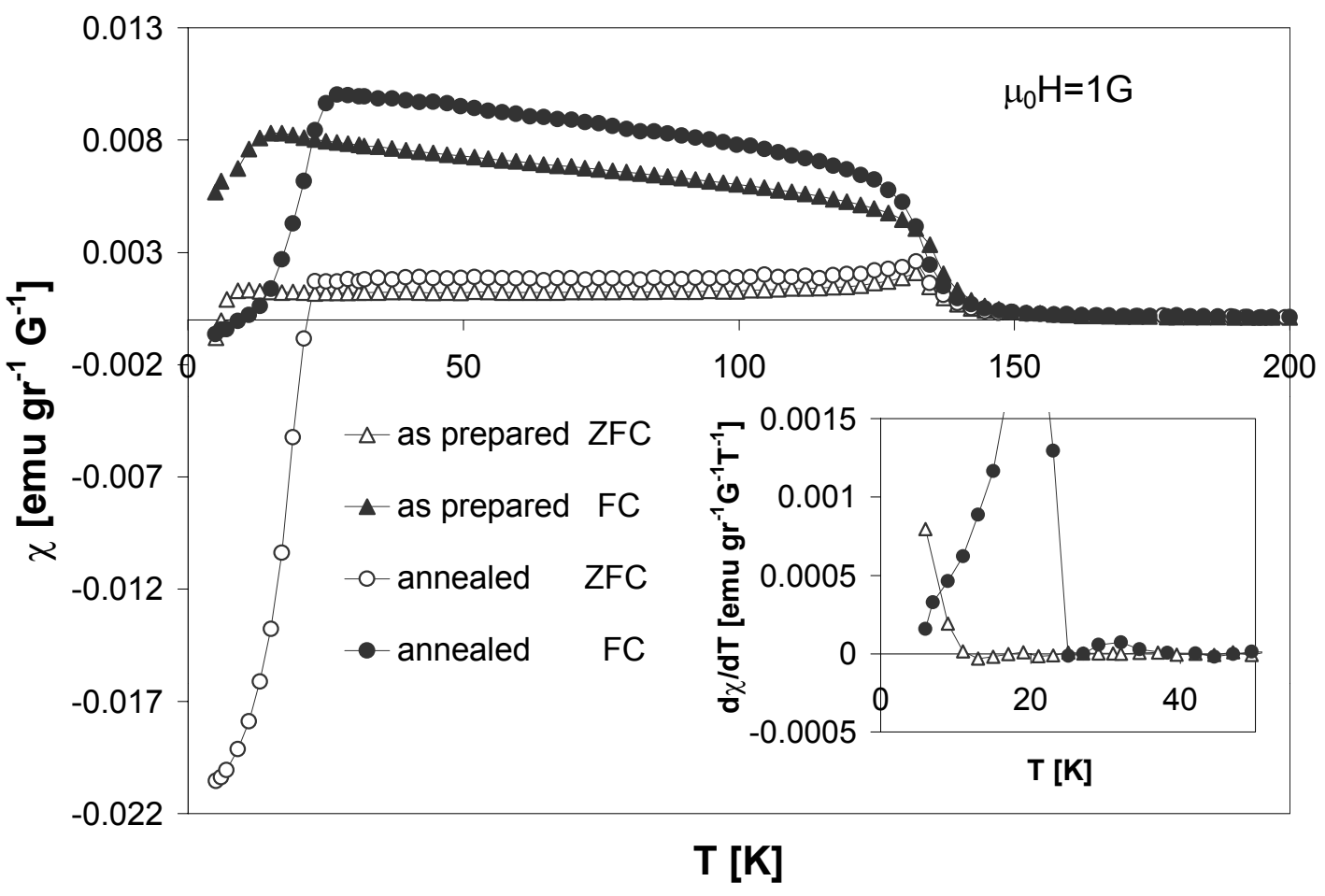

Figure 2

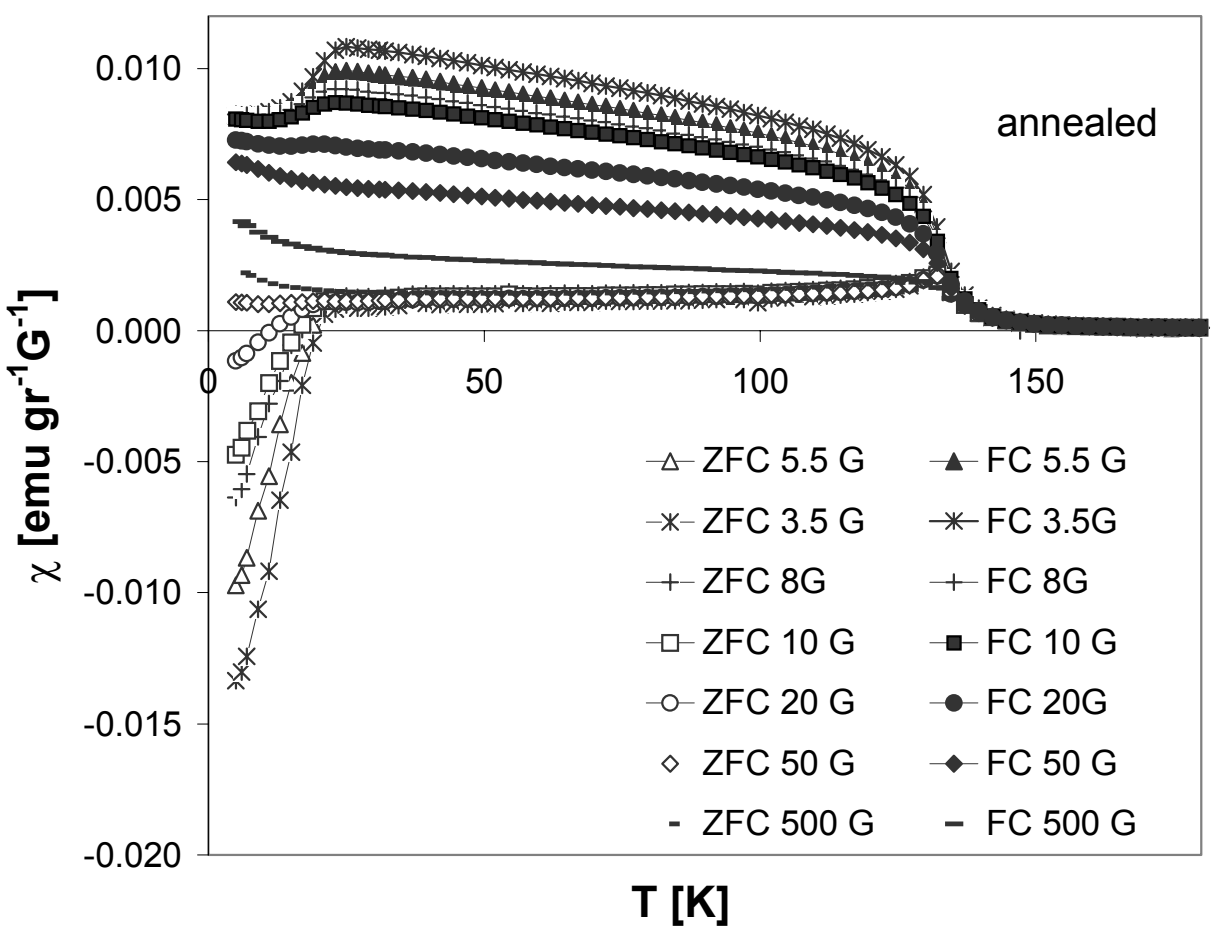

Figure 3

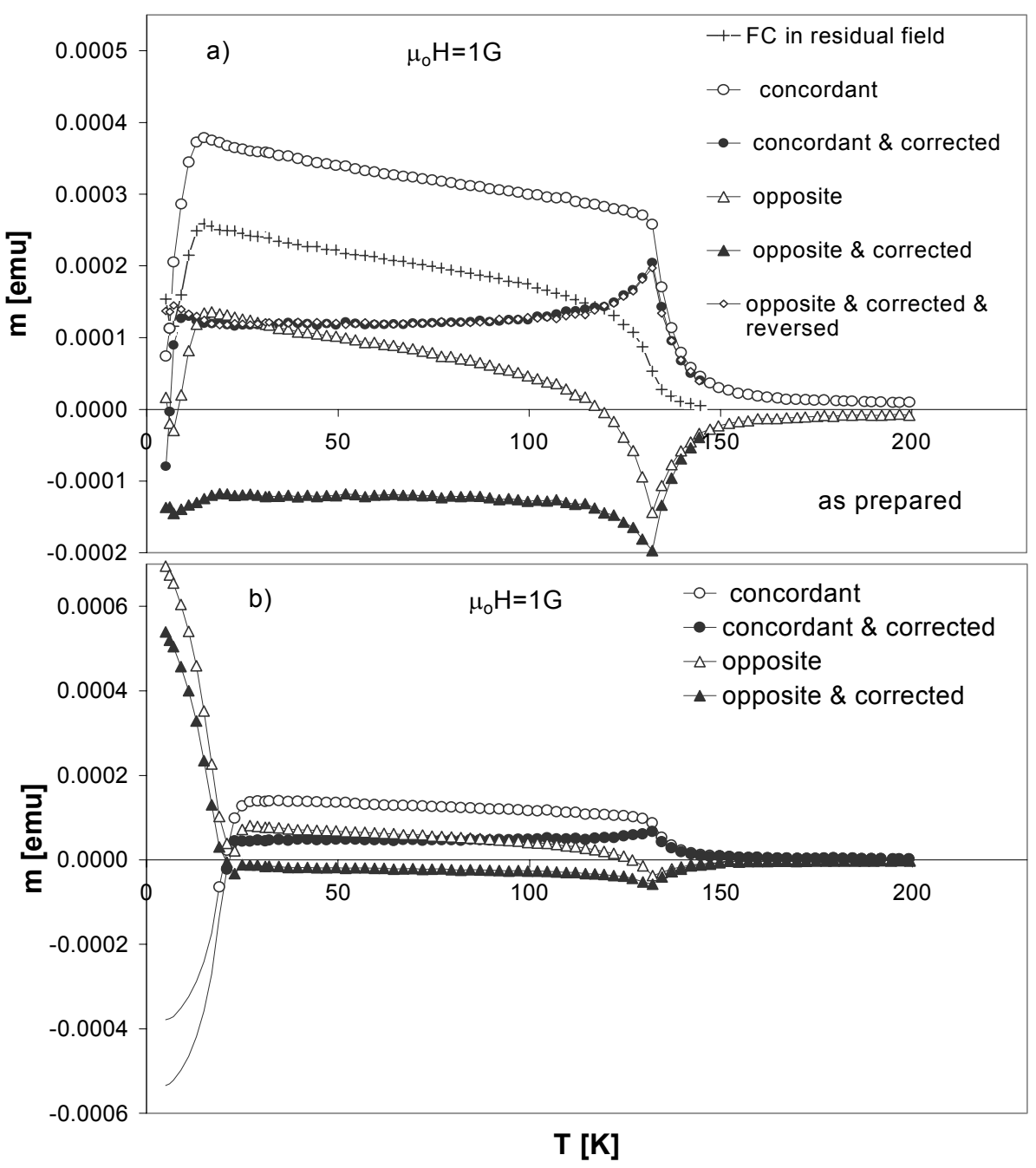

Figure 4

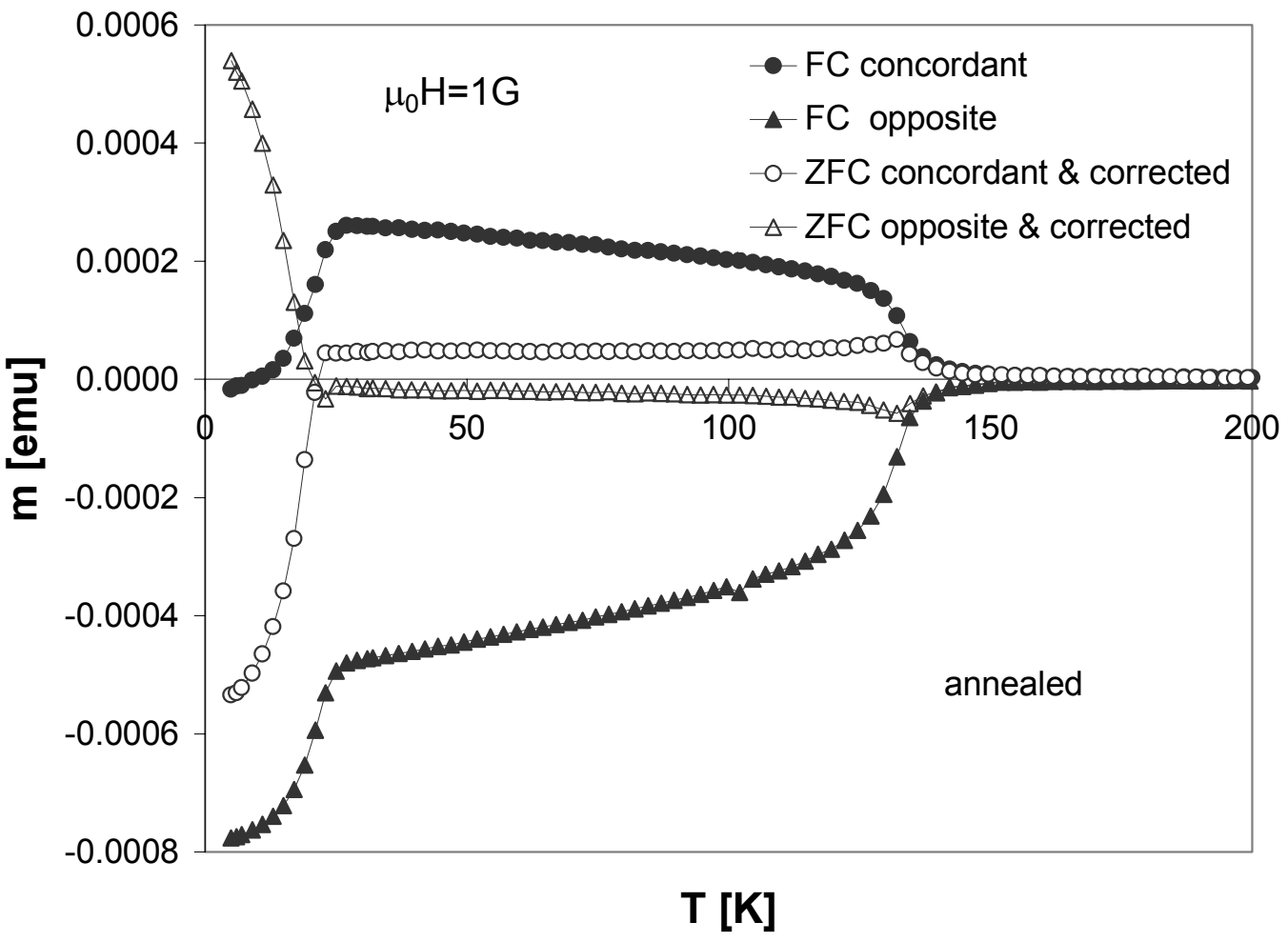

Figure 5

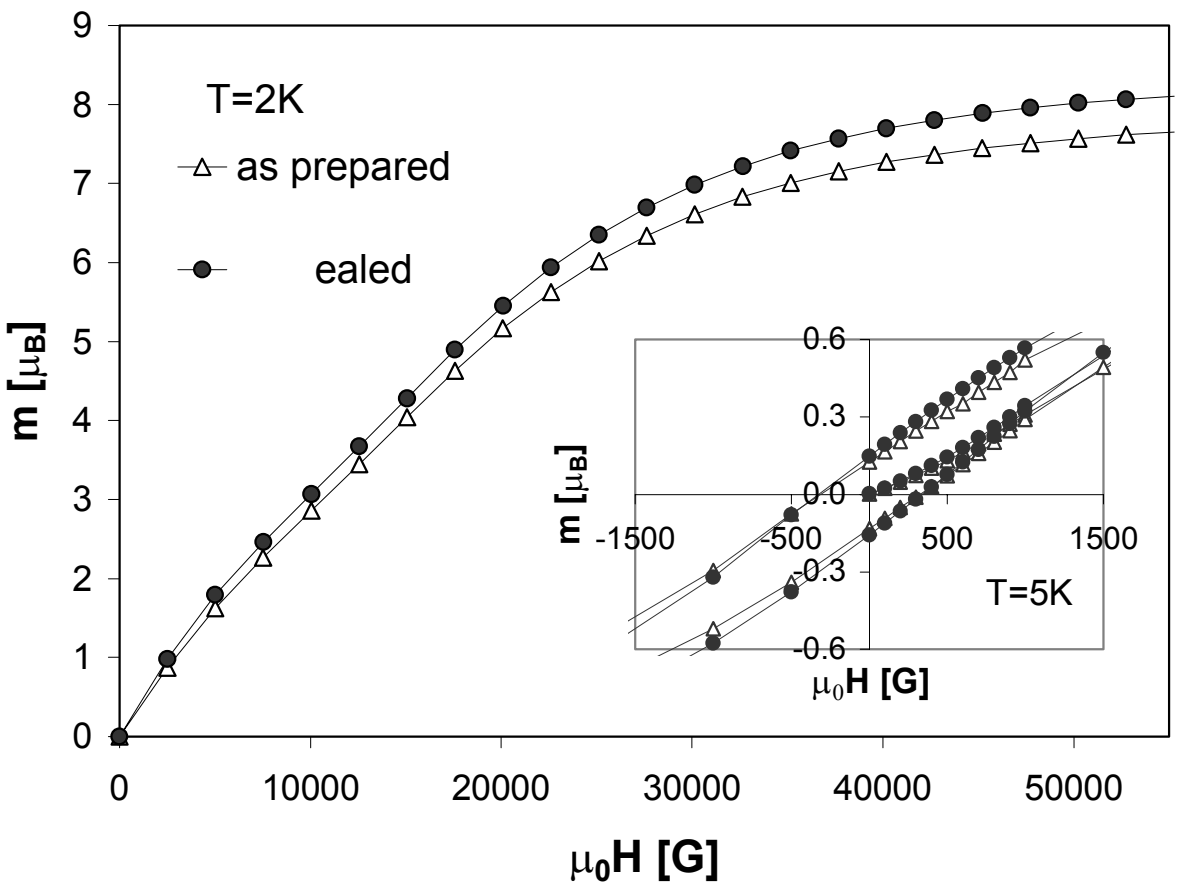

Figure 6